\documentclass[%
 reprint,
superscriptaddress,
 amsmath,amssymb,
 aps,
 prl,
floatfix,
]{revtex4-2}

\usepackage{graphicx}
\usepackage{dcolumn}
\usepackage{bm}

\usepackage[colorlinks,
           urlcolor=blue,
           linkcolor=blue,
           anchorcolor=blue,
           citecolor=blue
           ]{hyperref}
\usepackage{float}
\usepackage[dvipsnames]{xcolor}
\usepackage{soul}
\usepackage{siunitx}

\begin{document}

\preprint{APS/123-QED}

\title{\textbf{ Is Thermal Conductivity of Graphene Divergent and Higher Than Diamond?}}


\author{Zherui Han}
 \affiliation{School of Mechanical Engineering and the Birck Nanotechnology Center,\\
Purdue University, West Lafayette, Indiana 47907-2088, USA}
\author{Xiulin Ruan}%
 \email{ruan@purdue.edu}
 \affiliation{School of Mechanical Engineering and the Birck Nanotechnology Center,\\
Purdue University, West Lafayette, Indiana 47907-2088, USA}

\date{\today}

\begin{abstract}
The thermal conductivity of monolayer graphene is an outstanding challenge with no consensus reached on its exact value and length convergence so far. We consider four-phonon scattering, phonon renormalization, and an exact solution to phonon Boltzmann transport equation (BTE) from first principles.  Using this computational formalism with unprecedented sampling grid, we show that  when four-phonon scattering is included the thermal conductivity is convergent with system size at a room temperature value of 1300~W/(m$\cdot$K), which is lower than that of diamond. On the contrary, considering three-phonon scattering only yields divergence with size due to the momentum-conserving normal processes of flexural phonons. 

\end{abstract}

\maketitle


Graphene is a subject of extensive research due to its unique properties and exotic transport phenomena observed~\cite{Novoselov2004Electric,Zhang2005QuantumHall,Geim2007RiseofGraphene,Nair2008FineStructure,Ghosh2008ExtremelyHigh,Bonaccorso2010photonics}. Among them, thermal conductivity ($\kappa$) of graphene, which is dominated by phonon transport, is one of the open questions that remains widely debated: its reported divergence with system size is controversial and its value higher than diamond and graphite is widely believed. Despite numerous experimental studies, the measured values at room temperature are scattered~\cite{Balandin2008Superior,Chen2011Raman,Faugeras2010Thermal,Lee2011Thermal,Xu2014Lengthdependent}. The difficulty in experiments is exemplified in the first measurement of graphene thermal conductivity using Raman thermometry~\cite{Balandin2008Superior}. Later, strong nonequilibrium phonon transport was revealed which causes additional challenges in extracting the thermal conductivity~\cite{nl2017nonequilibrium,Vallabhaneni2016RamanReliability}.
In parallel, theoretical works also report a large spread of $\kappa$ between 800 to 3500~W/(m$\cdot$K) in atomistic simulations~\cite{Lindsay2010Flexural,Qiu2012Reduction,Lindsay2014PhononFirstprinciples,Fugallo2014Thermal,Fan2017ThermalEMD,Feng2018Fourphonon}.

To resolve the thermal conductivity of graphene, one must investigate two important yet elusive questions. One, the out-of-plane lattice vibration in graphene represented as the flexural phonons (ZA)~\cite{Mariani2007Flexural} contributes significantly to $\kappa$ and its scattering was thought to be greatly suppressed by a selection rule for three-phonon (3ph) scattering~\cite{Lindsay2010Flexural}. But such selection rule does not apply to four-phonon (4ph) scattering~\cite{Feng2018Fourphonon}, a higher-order mechanism found to be important in other solids~\cite{Feng2016fourphonon,Feng2017fourphonon,Xia2020Highthroughput,Ravichandran2020Phonon}. Two, except for the out-of-plane vibrations, graphene resembles the two-dimensional (2D) nonlinear lattice that was originated from the seminal Fermi-Pasta-Ulam-Tsingou (FPUT) model~\cite{fermi1955FPUT}. Heat transport in such perfect low-dimensional systems is non-Fourier~\cite{Lepri2016Thermal} and $\kappa$ should diverge with system size as $\kappa_{\rm 2D} \propto \log(L)$~\cite{Basile2006Momentum,Wang2012Logarithmic}. It is a matter of debate whether real quasi-2D systems like graphene can recover diffusive regime when $L\to \infty$ and acquires a finite intrinsic thermal conductivity.

Most theoretical studies do concern the above two questions but consensus has not been reached yet. Lindsay \textit{et al.} employed the linearized phonon Boltzmann transport equation (BTE) and coupled it with empirical potential~\cite{Lindsay2010Flexural} or first principles~\cite{Lindsay2014PhononFirstprinciples} at 3ph level. They proved that the exact solution to BTE is necessary for graphene due to its strong momentum-conserving scattering (normal process) and the length-dependent thermal conductivity didn't show convergence up to 50~$\mu$m. Their finite-size results at room temperature ($\sim$ 3000~W/(m$\cdot$K)) are in accordance with another theoretical study at 3ph level~\cite{Fugallo2014Thermal} reporting length convergence at around 1~mm, but in this study~\cite{Fugallo2014Thermal} the sampling grid of the Brillouin Zone was not checked for $\kappa$ convergence. Along this BTE approach, Feng and Ruan~\cite{Feng2018Fourphonon} computed 4ph scattering rates with empirical potential and showed strong 4ph effect in ZA mode and consequently, a great reduction in $\kappa$ to around 800~W/(m$\cdot$K). The sampling grid accessible then was relatively coarse. Gu \textit{et al.} further incorporated phonon renormalization into this calculation with empirical potential and argued that ZA scattering rates are reduced by temperature modification~\cite{Gu2019Revisiting}. However, the empirical interatomic potentials used in these studies are not accurate representations of experiments and the exact role of four-phonon scattering and the value of $\kappa$ remains elusive. Another mainstream approach is molecular dynamics (MD) with two different flavors in equilibrium (EMD) and nonequilibrium (NEMD) treatments. Regardless of classical statistics and empirical potential, MD approach is expected to capture all orders of anharmonicity. However, the EMD simulations show finite $\kappa$ around 2000~W/(m$\cdot$K)~\cite{Qiu2012Reduction,Fan2017ThermalEMD} while the NEMD simulations show logarithm length dependence up to several microns~\cite{Xu2014Lengthdependent,Ray2019Heatfluctuations}. Interestingly, one MD study presents a saturation of $\kappa$ when $L$ is extended to 100~$\mu$m~\cite{Barbarino2015Intrinsic}.

\begin{figure*}[ht]
    \centering
    \includegraphics[width=6.8in]{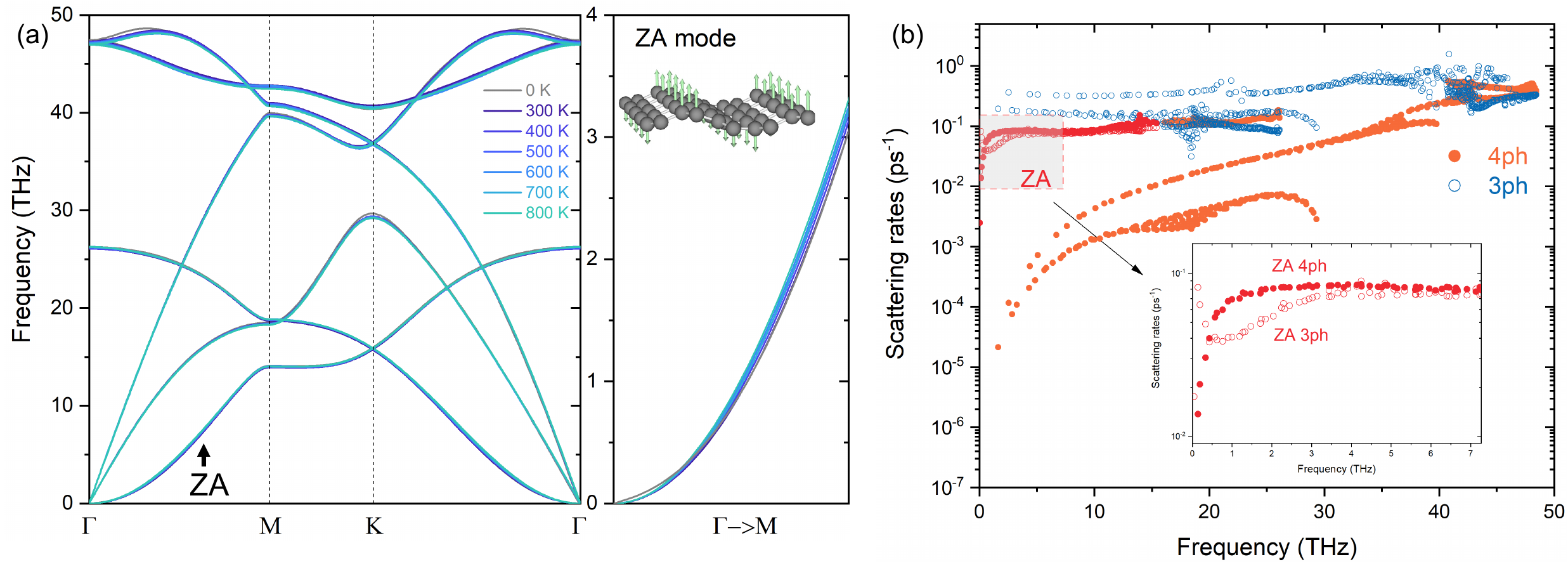}
    \caption{Phonon self-energy in graphene. (a) Phonon dispersion at finite temperatures by TDEP method. Dispersion at 0~K is the gound state calculation by density functional perturbation theory. The right panel is a zoom-in plot of ZA mode dispersion from $\Gamma$ to M in the Brillouin Zone. Schematic of ZA out-of-plane vibrations is created by an online visualization tool~\cite{Phonon}. (b) Spectral phonon scattering rates at 300~K. The scattering rate in $y-$axis is presented in logarithm scale. Three-phonon ($\tau_{\rm{3ph}}^{-1}$) and four-phonon scattering rates ($\tau_{\rm{4ph}}^{-1}$) are shown in hallow and filled circles, respectively. The scattering rates of ZA mode are marked in red and zoomed-in in the inset.}
    \label{disp-tau}
\end{figure*}

In this Letter, we revisit the thermal conductivity of graphene with first-principles-computed four-phonon scattering rates and the exact solution to BTE after careful check on convergence. In addition, to describe the temperature-dependent potential field we consider the phonon renormalization effect in both harmonic phonon dispersion~\cite{Hellman2011Lattice,Hellman2013TDEP} and anharmonic lattice dynamics~\cite{Hellman2013IFCs,Kim2018NuclearQuantum}. The considerations of four-phonon scattering and phonon renormalization have successfully explained the Raman linewidth of suspended graphene in our prior work~\cite{Han2022RamanLinewidth}. We note that the prediction in Ref.~\cite{Han2022RamanLinewidth} is supported by experimental measurements~\cite{Lin2011Anharmonic,nl2017nonequilibrium} which validate our approach to compute Raman-active phonon scattering rates, and in this work we extend the methodology to all phonon modes. The phonon BTE is exactly solved by an iterative scheme~\cite{Omini1995Iterative} incorporating both three-phonon and four-phonon scattering after we manage to store the iterative processes within one terabyte memory space accessible in modern supercomputer architecture. We compute from first principles that ZA mode has strong four-phonon scattering rates which are comparable to three-phonon counterpart. The sampling grid ($q-$mesh) is carefully checked for $\kappa$ convergence of infinitely long graphene sample and we see convergence for thermal conductivity including four-phonon scattering ($\kappa_{\rm 3ph+4ph}$) but not in thermal conductivity including only three-phonon scattering ($\kappa_{\rm 3ph}$). Finally, the temperature-dependent $\kappa$ is compared to diamond and found to be lower than diamond from 300~K to 800~K. At room temperature, we predict that graphene has an intrinsic thermal conductivity about 1300~W/(m$\cdot$K) and is finite. Our findings through rigorous first-principles calculations have implications in the thermal transport of quasi-low-dimensional systems, question the perception of graphene being a better heat conductor than diamond, and will motivate future experimental efforts.

We consider naturally occurring, monolayer graphene in our simulation with a vacuum space of 14 \r{A} between periodic graphene layers. The first-principles calculations are based on density functional theory as implemented in the \textsc{VASP} package~\cite{VASP1993}. The phonon renormalization effect is included by a temperature-dependent effective potential method (TDEP)~\cite{Hellman2013TDEP} to compute temperature-dependent phonon dispersions and interatomic force constants (IFCs)~\cite{Kim2018NuclearQuantum}. Phonon scattering rates summed up by Matthiessen's rule~\cite{ziman1960electronsphonons} $\tau_{\lambda}^{-1}=\tau_{\lambda,\rm{3ph}}^{-1}+\tau_{\lambda,\rm{4ph}}^{-1}+\tau_{\lambda,\rm iso}^{-1}$ are then computed by our \textsc{FourPhonon} code~\cite{han2021fourphonon}, which is an extension module to \textsc{ShengBTE} package~\cite{shengbte}. The exact solution of BTE is implemented in the same solver by a shared-memory parallel computing strategy~\cite{note}.

The computed phonon dispersion and phonon scattering rates are presented in Fig.~\ref{disp-tau}(a) and \ref{disp-tau}(b), respectively. Overall, graphene is quite rigid as shown by very small change of phonon frequency at finite temperatures (see Fig.~\ref{disp-tau}(a)). The frequency of in-plane optical phonons decreases with increasing temperature, a signature that has been analyzed and experimentally verified in our prior work~\cite{Han2022RamanLinewidth}. Contrary to the decreasing trend of frequency shift for in-plane optical phonons, we find that the flexural phonons are hardened with rising temperature (see Fig.~\ref{disp-tau}(a) right panel). This is a result of the coupling between flexural phonons and in-plane degrees of freedom in free-standing graphene~\cite{Mariani2007Flexural}. Renormalized flexural phonons do not have a strictly quadratic dispersion. With the renormalized phonon dispersion and temperature-dependent IFCs, one can then compute the spectral phonon scattering rates $\tau_{\lambda}^{-1}$ using Fermi's golden rule, as shown in Fig.~\ref{disp-tau}(b). Only the optical phonons and flexural phonons have comparable four-phonon scattering rates ($\tau_{\rm{4ph}}^{-1}$) with three-phonon counterpart ($\tau_{\rm{3ph}}^{-1}$), while the rest of phonon modes are dominated by three-phonon scattering. Optical phonons having large $\tau_{\rm{4ph}}^{-1}$ are understood as they easily satisfy the energy conservation in the recombination scattering events:  $\lambda_1+\lambda_2\to \lambda_3+\lambda_4$~\cite{Yang2020linewidth}. On a different ground, the reason for the flexural phonons lies in the selection rule for general quasi-2D systems~\cite{Lindsay2010Flexural} that only even numbers of flexural phonons can be involved in a phonon scattering event due to reflection symmetry. As a result, the phase space for three-phonon scattering is much smaller than that of four-phonon scattering~\cite{Feng2018Fourphonon}. Although the ZA mode was considered as the major heat carrier in graphene under the 3ph scattering picture~\cite{Lindsay2010Flexural,Lindsay2014PhononFirstprinciples,Seol2010SupportedGraphene}, this statement may no longer be true after its strong $\tau_{\rm{4ph}}^{-1}$ is included, which would greatly reduce the predicted thermal conductivity.

\begin{figure}[h]
    \centering
    \includegraphics[width=3.4in]{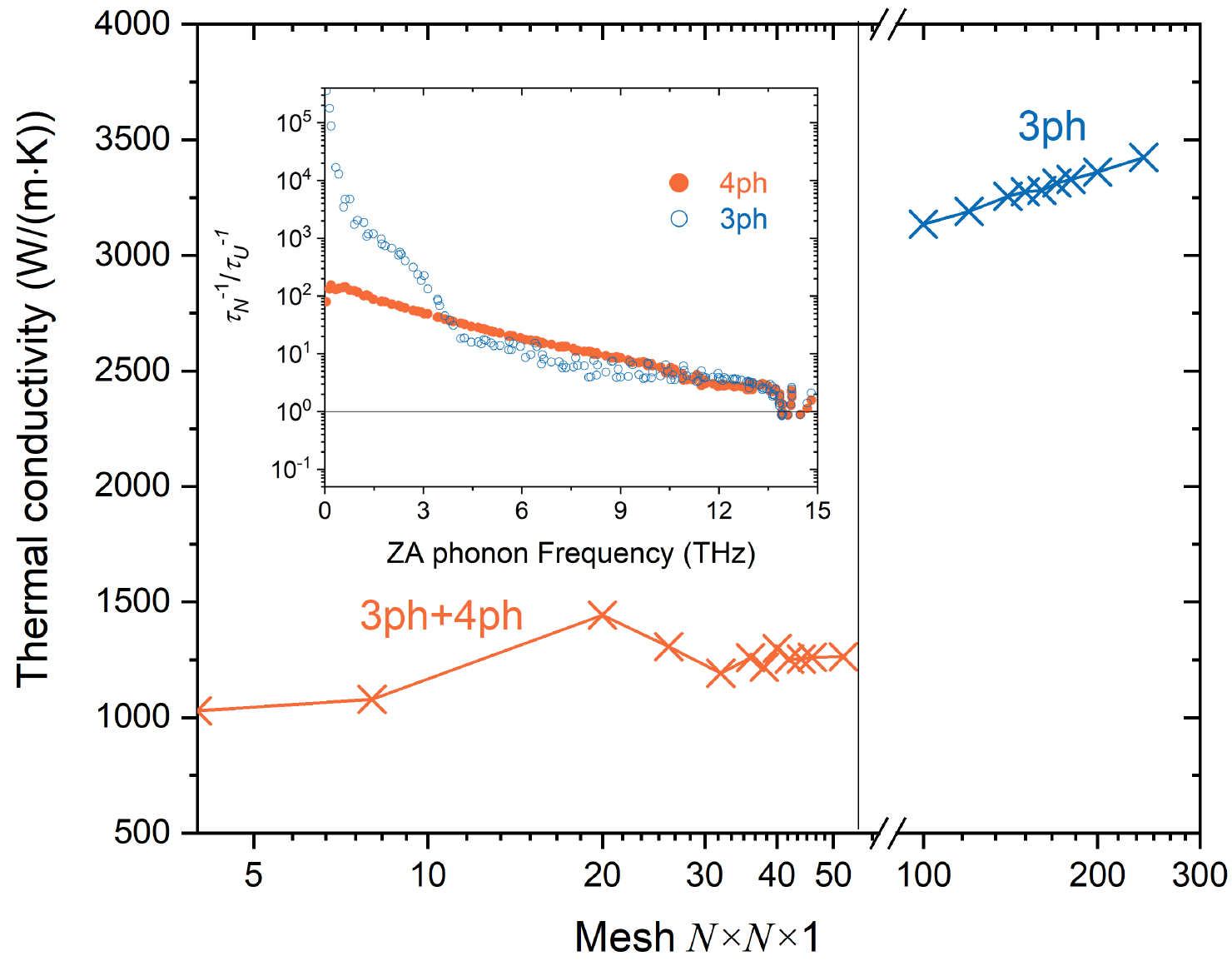}
    \caption{Convergence of thermal conductivity of graphene. Mesh size in $x-$axis is presented in logarithm scale. Results including 3ph only and 3ph+4ph together are shown in blue and orange markers, respectively. The inset shows the ratio of ZA mode scattering rates from normal ($\tau^{-1}_N$) and Umklapp processes ($\tau^{-1}_U$) for both 3ph (hallow blue dots) and 4ph channels (filled orange dots). The ratio in the inset has logarithm scale.}
    \label{convergenceTC}
\end{figure}

We next solve for the intrinsic thermal conductivity by an iterative scheme~\cite{Omini1995Iterative} to account for the collective phonon excitations. This approach distinguishes the normal processes (N) that are momentum-conserving and Umklapp processes (U) that are resistive. Such a treatment is important for graphene since it has strong normal scattering process associated with its phonon hydrodynamics nature~\cite{Lee2015Hydrodynamic,Cepellotti2015Hydrodynamics}. The coupled equations are presented in the Supplymental Material~\cite{supply}. For the BTE solution to generate meaningful results, one has to check the convergence of $\kappa$ with respect to the sampling grid in the Brillouin Zone ($q-$mesh). A convergence being reached means that an infinitely large system has finite thermal conductivity. Figure~\ref{convergenceTC} shows our test of convergence when 3ph or 3ph+4ph is included in the calculations. Mesh size $N\times N\times1$ is uniform in-plane. In all these calculations, the energy broadening factor is unity to ensure the accuracy within our computational power. We find that with four-phonon scattering, $\kappa_{\rm 3ph+4ph}$ converges around a mesh size of $N=40$ and further increasing the mesh to $N=52$ does not change the value of $\kappa$. In contrast, $\kappa_{\rm 3ph}$ does not reach convergence up to $N=240$ and scales logarithmly with mesh size $N$. Note that the prior work~\cite{Fugallo2014Thermal} used $N=128$ for 3ph. Our numerical results share similarities with a theoretical work on carbon nanotube (CNT)~\cite{Mingo2005Length} where the authors estimated that third-order anharmonicty leads to a divergence of thermal conductivity due to vanishing scattering of long wavelength phonons. They empirically showed that higher order process can remove such divergence. While the study on CNT is worth further investigations~\cite{Barbalinardo2021UltrahighCNT,Bruns2022Comment,Barbalinardo2022Reply}, we show here that four-phonon scattering converges the thermal conductivity of graphene from first principles. To inspect the origin of this behavior, we decompose the N/U scattering for ZA mode and plot the ratio of two types of scattering events $\tau^{-1}_N/\tau^{-1}_U$ in the inset of Fig.~\ref{convergenceTC}. Apart from the dominant role of normal process for both scattering channels, this plot indicates that for 3ph the N/U ratio keeps rising for low-frequency flexural phonons. Given the fact that $\tau^{-1}_N$ decreases when $\omega\to 0$, we conclude that 3ph of long wavelength flexural phonons is almost entirely contributed by normal processes. The extreme case is the heat conduction in 2D nonlinear lattice where all scattering processes are momentum-conserving~\cite{Wang2012Logarithmic} and the thermal conductivity is logarithmic divergent. Thus, we explain the finite intrinsic thermal conductivity of graphene from two related arguments: one, four-phonon scattering provides an additional scattering channel; two, for long wavelength phonons, four-phonon scattering has considerable number of resistive U scattering events but three-phonon processes are nearly all N scattering events implying a divergence.

\begin{figure}[h]
    \centering
    \includegraphics[width=2.7in]{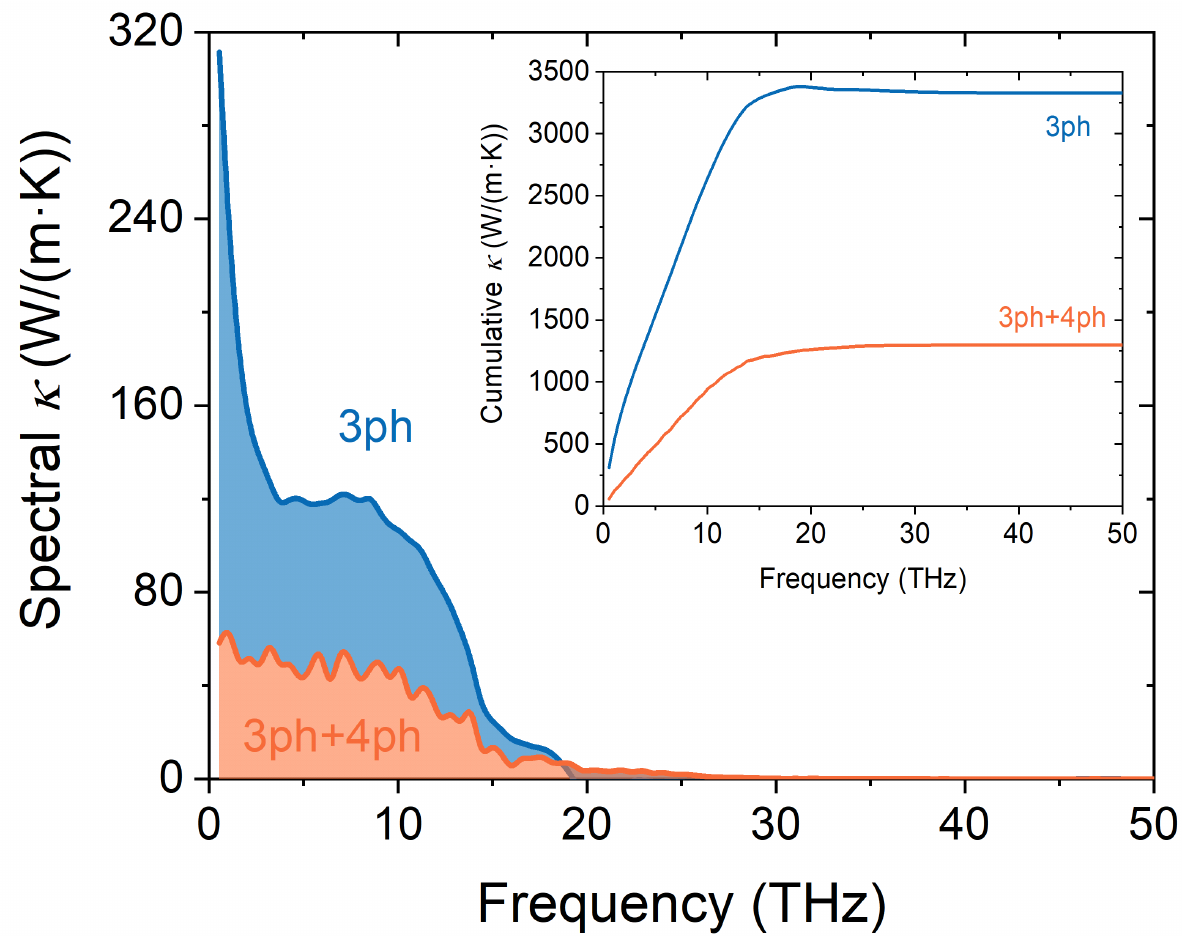}
    \caption{Spectral contributions to the thermal conductivity of graphene at room temperature without boundary scattering. The inset shows the cumulative thermal conductivity as a function of phonon frequency. The 3ph case presented here is calculated at $N=180$ and note that it is not converged with $N$.}
    \label{spectralTC}
\end{figure}

The above arguments can further be seen in the spectral $\kappa$ as shown in Fig.~\ref{spectralTC}. The 3ph and 3ph+4ph pictures have distinct spectral trend for low-frequency phonons. While the 3ph+4ph calculation can saturate the spectral $\kappa$ when $\omega\to 0$, the 3ph only calculation shows a up-soaring spectral trend. The integration of the shaded area in Fig.~\ref{spectralTC} should give the total $\kappa$ and this spectral analysis indicates that the 3ph case has a singularity at near-zero frequency.

\begin{figure}[h]
    \centering
    \includegraphics[width=3.5in]{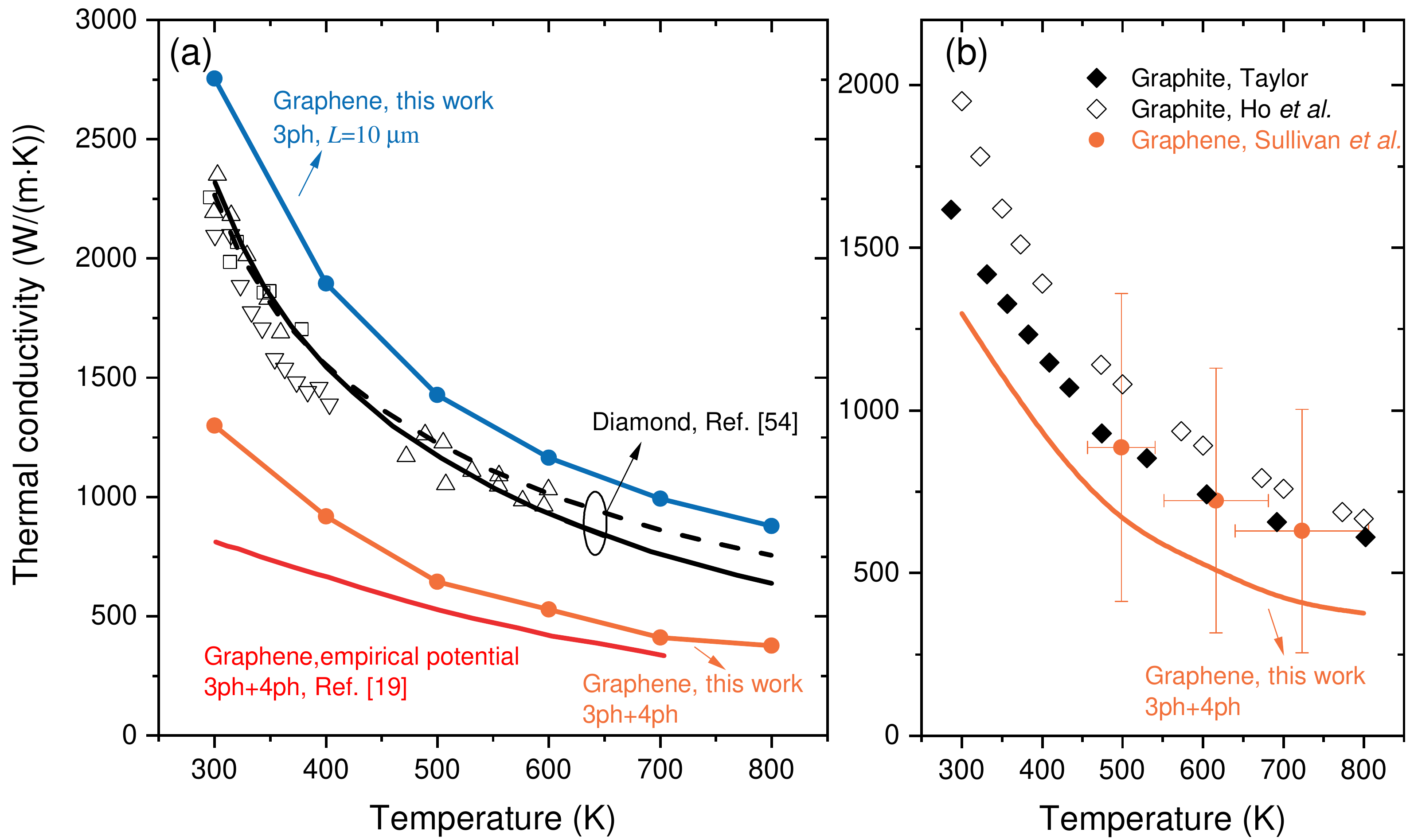}
    \caption{Thermal conductivity of graphene as a function of temperature. (a) The prior theoretical work on graphene with empirical potential and four-phonon scattering~\cite{Feng2018Fourphonon} is presented in red solid line. Our results are presented in connected solid lines with orange being results from 3ph+4ph, blue being results from 3ph scattering only and a boundary scattering of $L=10~\mu$m. Computational results for diamond~\cite{Ravichandran2018Unified} are presented in black lines (3ph results in dashed line and 3ph+4ph results in solid line). Several experiments on diamond are also plotted here (squares~\cite{Berman1975Diamond}, triangles~\cite{Olson1993Diamond}, inverted triangles~\cite{Onn1992Diamond}). (b) Our results compared to a recent Raman measurement on suspended graphene~\cite{nl2017nonequilibrium} (filled orange circles) and an experiment on graphite~\cite{Taylor1966Graphite,Ho1974Thermal} (black filled diamonds).}
    \label{TC}
\end{figure}

Finally, we present the calculated thermal conductivity of graphene as a function of temperature and some comparisons to other carbon allotropes in Fig.~\ref{TC}. Since our 3ph calculations are not converged within the accessible mesh size, we consider a $10~\mu$m diameter graphene sample and add a boundary scattering term in our simulation~\cite{Lindsay2014PhononFirstprinciples,Fugallo2014Thermal}: $\tau^{-1}_{\lambda,b}=\left|\mathbf{v}_\lambda\right| / L$, where $\mathbf{v}_\lambda$ is the group velocity of phonon mode $\lambda$. Our results on finite-size $\kappa_L$ at 3ph level are consistent with prior first-principles studies~\cite{Lindsay2014PhononFirstprinciples,Fugallo2014Thermal}. The comparison in Fig.~\ref{TC}(a) shows that incorporating 4ph in the calculation reduces $\kappa_{\rm 3ph}$ nearly by half. Nevertheless, our results of $\kappa_{\rm 3ph+4ph}$ are higher than previous simulation with empirical potential at moderate mesh size~\cite{Feng2018Fourphonon}. At room temperature, we predict that $\kappa_{\rm 3ph+4ph}=1298$~W/(m$\cdot$K), and the thermal conductivity of graphene is lower than that of diamond from 300~K to 800~K. The first-principles results for diamond (black lines in Fig.~\ref{TC}(a)) are from Ref.~\cite{Ravichandran2018Unified} where similar methodologies were applied including 3ph/4ph scattering, phonon renormalization and iterative solution to BTE. The dashed black line is their result at 3ph level while solid black line at 3ph+4ph level. Note that the thermal conductivity of diamond is well received, with good agreements between simulations and experiments~\cite{Berman1975Diamond,Onn1992Diamond,Wei1993TCDiamond,Olson1993Diamond}. Contrary to the situation of diamond, no consensus has been reached on the measured thermal conductivity value of graphene so far. While our finding challenges a popular perception that graphene is a better heat conductor than diamond, the numerical results are self-consistent with advancements in theories and computational power. Under current computational formalism, a plausible reason for this observation is that diamond does not have strong four-phonon scattering for acoustic modes~\cite{Feng2017fourphonon,Ravichandran2018Unified} but flexural phonons in graphene do have large four-phonon scattering rates originated from its 2D nature. In this sense, the reduction of $\kappa$ from diamond to graphene is understandable. Another consequence of our results is that the predicted graphene thermal conductivity is lower than experimentally reported $\kappa$ of graphite~\cite{Taylor1966Graphite,Ho1974Thermal} (Fig.~\ref{TC}(b)). Theoretical study on graphite considering both four-phonon scattering and phonon renormalization is not seen in literature yet. Future studies could be focusing on the layer-dependent transport behavior of multilayer graphene and graphite. In Fig.~\ref{TC}(b), we also cautiously compare our results to a recent Raman measurement on suspended graphene sample~\cite{nl2017nonequilibrium} that is based on apparent phonon temperature. We expect such comparison may motivate further experimental efforts to fully resolve the intriguing finding here. A tentative explanation might be that in bulk phase graphite the four-phonon scattering of ZA mode is suppressed by interlayer interactions.

In summary, we conduct a first-principles study on the thermal conductivity of monolayer graphene. Our calculations include four-phonon scattering, phonon renormalization effect and the exact solution to BTE, all of which are state-of-the-art computational formalism. Our methods can reveal both phonon properties, their detailed scatterings and eventually the thermal conductivity of graphene over a wide range of temperature. Part of the computed results have been supported by recent experiments. Our results provide a strong computational evidence of $\kappa$ convergence to date. We expect that our study may inspire further experimental efforts on graphene and theoretical understanding on general low-dimensional systems.

\begin{acknowledgments}
Detailed formalism and methods, along with length-dependent $\kappa$ and comparison to prior studies are presented in the Supplymental Material~\cite{supply}. We are grateful to the insightful discussions with Professor Li Shi at the University of Texas at Austin. X. R. and Z. H. were supported by the U.S. National Science Foundation (No.~2015946). Simulations were performed at the Rosen Center for Advanced Computing (RCAC) of Purdue University. 

\end{acknowledgments}


\bibliography{Reference}

\begin{thebibliography}{60}
\bibitem [{not()}]{note}%
  \BibitemOpen
  \href@noop {} {}\bibinfo {note} {Incorporating four-phonon scattering to the
  iterative solver is nontrivial since the scattering matrix and phonon
  information consumes great amount of storage. For future usage and community
  adoption, we use OpenMP parallelism to store the scattering matrix in memory.
  The memory consumption scales with the mesh size $N$ to the fifth power. The
  largest mesh size case can be done on a flat node equipped with 1 TB
  memory.}\BibitemShut {Stop}%
\end{thebibliography}

\end{document}